\documentclass[conference]{IEEEtran}
\usepackage{amsmath,amssymb}
\usepackage{graphicx}
\usepackage{mathrsfs}
\usepackage{tikz}
\usepackage{booktabs}
\usepackage{floatflt}
\usepackage{enumerate}
\usepackage{psfrag}	
\usepackage{array}
\usepackage{multirow,hhline}
\usepackage{exscale}
\usepackage{color}
\usepackage{colortbl}
\usepackage{epsfig,subfigure}
\usepackage{cite}
\usepackage{amsthm}

\newtheorem{theorem}{Theorem}

\newtheorem{definition}{Definition}

\begin{document}

\IEEEoverridecommandlockouts
\title{Lattice Coding and the Generalized Degrees of Freedom of the Interference Channel with Relay}
\author{
\IEEEauthorblockN{Anas Chaaban and Aydin Sezgin}
\IEEEauthorblockA{Chair of Communication Systems\\
RUB, 44780 Bochum, Germany\\
Email: {anas.chaaban@rub.de, aydin.sezgin@rub.de}}
}

\maketitle


\begin{abstract}
The generalized degrees of freedom (GDoF) of the symmetric two-user Gaussian interference relay channel (IRC) is studied. While it is known that the relay does not increase the DoF of the IC, this is not known for the more general GDoF. For the characterization of the GDoF, new sum-capacity upper bounds and lower bounds are derived. The lower bounds are obtained by a new scheme, which is based on functional decode-and-forward (FDF). The GDoF is characterized for the regime in which the source-relay link is weaker than the interference link, which constitutes half the overall space of channel parameters. It is shown that the relay can indeed increase the GDoF of the IRC and that it is achieved by FDF.
\end{abstract}

\section{Introduction}
The exact characterization of the capacity of interference networks is an open problem for several decades now. Given the difficulty of the problem, there is shift of paradigm to provide an approximate characterization of the capacity, referred to as the degrees of freedom (DoF), which gets asymptotically tight for high signal-to-noise power ratios ($\mathsf{SNR}$). While the DoF provides interesting insights into the behaviour of the system, the so-called generalized degrees of freedom, or GDoF~\cite{EtkinTseWang}, is a much more powerful metric, as it allows different signal strengths and thus captures a large variety of scenarios.

The setups gets even more interesting for cases, in which some of the nodes are dedicated relays. It is known that relaying in wireless networks can play a vital role in improving its performance in terms of coverage and achievable rates. As such, a relay can help the network by establishing cooperation between the nodes in the network. Interestingly enough, the capacity of even the basic point-to-point (P2P) relay channel~\cite{CoverElgamal} (without interference) is an open problem, although there exist good approximations of the capacity of the Gaussian P2P relay channel within one bit \cite{AvestimehrDiggaviTse}.

The improvements obtained depend heavily on the capability of and restrictions at the relay, such as cognition and causality.
For example, for a network with two transmitters, two receivers, and a relay referred to as the interference relay channel (IRC), several capacity bounds have been derived with a causal or with a cognitive relay \cite{MaricDaboraGoldsmith,SahinErkip,SahinErkip_Cognitive,SridharanVishwanathJafarShamai,TianYener_PotentRelayJournal,RiniTuninettiDevroye_ISIT,RiniTuninettiDevroye_ITW}. As for the relay and the interference channel (IC) individually, the capacity of the IRC remains an open problem.

Surprisingly, in characterizing the gains in terms of DoF by deploying a relay in a wireless interference network, it was shown in~\cite{CadambeJafar_ImpactOfRelays} that relaying does not increase the DoF. As a consequence of this result, the DoF of the IRC is the same as that of the IC, i.e., DoF=1.
One question which immediately arises with this result is whether this is also true in terms of GDoF.

In this paper, we investigate the GDoF for the symmetric Gaussian IRC. Shortly, our contribution includes deriving new upper bounds on the sum-capacity, providing new achievable sum-rates by proposing a ``functional decode-and-forward''~\cite{OngKellettJohnson} (FDF) scheme. The distinct feature of the achievable strategy is that the overall message is split in three parts, namely a private, a common and a cooperative public part. While the former two are in use already in the basic IC, the latter one, encoded using nested lattices, is of particular value to overcome the multiple-access-bottleneck at the relay. We characterize the GDoF of the IRC for all cases in which the interference link is stronger than the source-relay link. This characterized regime covers half the space of all possible channel parameters for the IRC, and is especially interesting for the IRC with weak interference. It turns out that while a relay does not increase the DoF of the IC, it does increase its GDoF. In the next section, we formally define the IRC and the notation used in the paper.

\section{Network Model and Notation}
\label{Section:Model}

\begin{figure}[t]
\centering{
\includegraphics[width=.9\columnwidth]{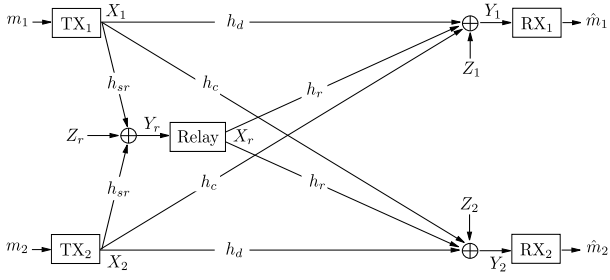}
}
\caption{The 2-user interference relay channel (IRC).}
\label{2UserICFDR}
\end{figure}

In the symmetric Gaussian IRC (Fig. \ref{2UserICFDR}), transmitter $i$, $i\in\{1,2\}$, has a message $m_i$ uniformly distributed over the set $\mathcal{M}_i\triangleq\{1,\dots,2^{nR_i}\}$, to be sent to receiver $i$. The message $m_i$ is encoded into an $n$-symbol codeword $X_i^n$, where $X_{ik}$ is a real valued random variable, and transmits this codeword. At time instant $k$, the input-output equations of this setup are given by
\begin{align*}
y_{ik}&=h_{d}x_{ik}+h_{c}x_{jk}+h_{r}x_{rk}+z_{ik},\\
y_{rk}&=h_{sr}x_{1k}+h_{sr}x_{2k}+z_{rk}.
\end{align*}                                                                                                                                                                                                                                                                                                                                     
for $i\neq j$, $i,j\in\{1,2\}$. The coefficients $h_{d},h_{c},h_r,h_{sr}\geq0$ are real valued channel gains, and $x_{rk}$ is the transmit signal at the relay at time instant $k$. The relay is causal, which means that $x_{rk}$ is a function of the previous observations at the relay, i.e., $x_{rk}=f_r(y_r^{k-1})$. The source and relay signals must satisfy a power constraint $\mathbb{E}[X_i^2]\leq P$, $i\in\{1,2,r\}$. The receivers' additive noise $z_1$, $z_2$, and $z_r$ is Gaussian with zero mean and unit variance.

After receiving $y_i^n$, receiver $i$ uses a decoder to detect $m_i$ by processing $y_i^n$. The messages set, encoders, and decoders define a code denoted $(2^{nR_1},2^{nR_2},n)$, with an error probability $P_e$ defined by $P_e=P\left(\hat{m}_1\neq m_1 \text{ or } \hat{m}_2\neq m_2\right)$. A rate pair $(R_1,R_2)$ is said to be achievable if there exists a sequence of $(2^{nR_1},2^{nR_2},n)$ codes such that $P_e\to0$ as $n\to\infty$. The capacity region $\mathcal{C}$ of the IRC is defined as the closure of the set of these achievable rate pairs, and the sum-capacity $C_\Sigma$ is the maximum achievable sum-rate $R_\Sigma=R_1+R_2$, i.e., $C_\Sigma=\max_{(R_1,R_2)\in\mathcal{C}}R_\Sigma$. The GDoF of the IRC is defined as follows.
\begin{definition}
Let the following variables represent the channel strength (as in \cite{EtkinTseWang})
\begin{align}
\label{ABGDef}
\alpha=\frac{\log(h_c^2P)}{\log(h_d^2P)},\quad \beta=\frac{\log(h_r^2P)}{\log(h_d^2P)},\quad\gamma=\frac{\log(h_{sr}^2P)}{\log(h_d^2P)}.
\end{align}
We define the GDoF $d(\alpha,\beta,\gamma)$ or simply $d$ as
\begin{align}
\label{GDoFDef}
d=\lim_{\substack{P\to\infty}}\frac{C_\Sigma(\alpha,\beta,\gamma)}{\frac{1}{2}\log(h_d^2P)}.
\end{align}
\end{definition}

Throughout the paper, we use $x^n$ to denote the length-$n$ sequence $(x_1,\dots,x_n)$, and we use $C(x)=(1/2)\log(1+x)$, $C^+(x)=\max\left\{0,C(x)\right\}$.

\section{Main Result}
\label{Section:Summary}

The main statement of the paper is characterizing the GDoF of the IRC for all cases where $h_{sr}^2\leq h_c^2$. The GDoF in this case is given in the following theorem.

\begin{theorem}
\label{Theorem:GDoF}
The GDoF of the IRC with $\gamma\leq\alpha$ is given by
\begin{align}
\label{GDoFThm}
d=\min\left\{\begin{array}{c}
2\max\{1,\beta\}\\
2\max\{1,\gamma\}\\
\max\{1,\alpha,\beta\}+\max\{1,\alpha\}-\alpha\\
2\max\{1,\alpha\}+\gamma-\alpha\\
2\max\{\alpha,\beta,1-\alpha\}\\
2\max\{\alpha,1+\gamma-\alpha\}\end{array}\right\}.
\end{align}
\end{theorem}

\begin{figure}[t]
\centering
\psfragscanon
\psfrag{a}[t]{$\alpha$}
\psfrag{d}[b]{$d$}
\psfrag{g}[l]{\footnotesize$\gamma$}
\psfrag{b}[l]{\footnotesize $\beta$}
\psfrag{z}[l]{\footnotesize $0$}
\psfrag{o}[r]{\footnotesize$1$}
\psfrag{t}[r]{\footnotesize$2$}
\psfrag{1pbma}[l]{\footnotesize $1+\beta-\alpha$}
\psfrag{2p2gm2a}[lb]{\footnotesize $2+2\gamma-2\alpha$}
\psfrag{2a}[bl]{\footnotesize$2\alpha$}
\includegraphics[width=0.9\columnwidth]{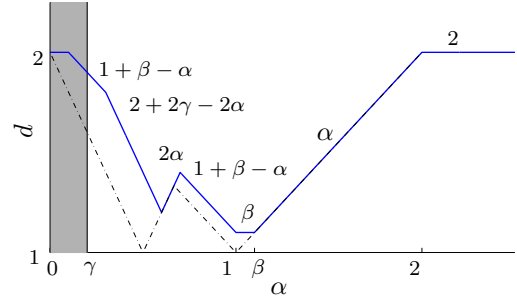}
\caption{$\beta=1.1$ and $\gamma=0.2$. A case where the GDoF of the IRC is larger than that of the IC in both the weak and the strong interference regimes.}
\label{Fig:GDoFeg1102}
\end{figure}

\begin{figure}[t]
\centering
\psfragscanon
\psfrag{a}[t]{$\alpha$}
\psfrag{d}[b]{$d$}
\psfrag{g}[l]{\footnotesize$\gamma$}
\psfrag{b}[l]{\footnotesize $\beta$}
\psfrag{z}[l]{\footnotesize $0$}
\psfrag{o}[r]{\footnotesize$1$}
\psfrag{t}[r]{\footnotesize$2$}
\psfrag{2g}[]{\footnotesize$2\gamma$}
\includegraphics[width=0.9\columnwidth]{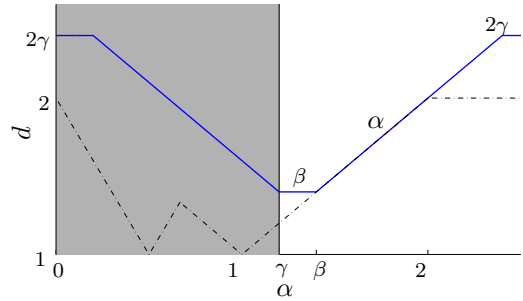}
\caption{$\beta=1.4$ and $\gamma=1.2$. A case where the GDoF of the IRC is larger than that of the IC in both the strong and the very strong interference regimes.}
\label{Fig:GDoFeg1412}
\end{figure}

Figures \ref{Fig:GDoFeg1102} and \ref{Fig:GDoFeg1412} show the GDoF for two examples of the IRC. The GDoF of the IC is also shown (dash-dotted) for comparison. The characterization of the GDoF in the shaded area, where $\alpha<\gamma$, is not considered in this paper. The proof of this theorem is provided in the next section. Namely, in Section \ref{Section:UpperBounds} we provide the sum-capacity upper bounds that translate to this GDoF, in Section \ref{Section:FDF} we describe the scheme that is used to achieve it, and in Section \ref{Section:GDoF} we sketch the proof of Theorem \ref{Theorem:GDoF}.

\section{Upper Bounds}
\label{Section:UpperBounds}
We start by providing the following bounds that can be obtained from the cut-set bounds \cite{CoverThomas} applied to the IRC. The cut-set bounds for the IRC are expressed in \cite{MaricDaboraGoldsmith,ChaabanSezgin_Asilomar}.

\begin{theorem}
\label{CutSetBound}
$C_\Sigma$ is upper bounded by \\$C_\Sigma\leq 2C((|h_d|+|h_r|)^2P)$ and $C_\Sigma\leq 2C(h_d^2P+h_{sr}^2P)$.
\end{theorem}
Using the definition of $\beta$ in \eqref{ABGDef} and the first bound in Theorem \ref{CutSetBound}, we can write $C_\Sigma\leq 2C((|h_d|+|h_r|)^2P)\leq 2\max\{C(h_d^2P),C((h_d^2P)^{\beta})+2$, which, by using \eqref{GDoFDef} translates to the first argument in the $\min$ in \eqref{GDoFThm}. The second argument in \eqref{GDoFThm} can be obtained similarly from the second bound in Theorem \ref{CutSetBound}. Using a similar method, the third and fifth arguments in \eqref{GDoFThm} can be obtained from the bounds in \cite[Theorems 1 and 2]{ChaabanSezgin_Asilomar}. The remaining expressions in \eqref{GDoFThm} are obtained from the following theorem.

\begin{theorem}
\label{SumBound2}
$C_\Sigma$ is upper bounded by
\begin{align}
\label{Bound1}
C_\Sigma&\leq C(2h_{sr}^2P)+C(h_d^2P+h_c^2P)+C^+\left(h_d^2/h_c^2-1\right)\\
\label{Bound2}
C_\Sigma&\leq 2C (h_c^2/h_{sr}^2+\left(1-h_d/h_c\right)^2 )+2C(2h_{sr}^2P).
\end{align}
\end{theorem}

Due to space limitations, we only provide a sketch of the proof of \eqref{Bound1} in Appendix \ref{Proof:SumBound2}. In the next section, we provide a GDoF achieving scheme for the IRC.

\section{Achievability: Functional Decode-and-Forward}
\label{Section:FDF}
Using nested-lattice coding and lattice alignment, we establish a cooperation strategy between the relay and the users. This scheme is denoted ``Functional Decode-and-Forward'' (FDF) using the terminology of \cite{OngKellettJohnson}. We use three kinds of messages in FDF, private (P), common (C), and cooperative public (CP) messages. The private and the common messages are the same as those used by Etkin et al. in the IC \cite{EtkinTseWang}. The CP message itself is also split into $K$ sub-messages. The superposition of the CP messages is decoded by the relay, and forwarded to the destinations. Using backward decoding, the sum-rates given in the following theorems are achievable.

\begin{theorem}
\label{FDFWI}
The sum-rate $R_\Sigma= 2(R_p+R_c+R_{cp})$ is achievable where
\begin{align}
\label{PRateConstraint}
R_{p}&\leq C\left(\frac{h_d^2P_p}{1+h_c^2P_p}\right),\\
\label{CRateConstraint1}
R_{c}&\leq C\left(\frac{\min\{h_d^2,h_c^2\}P_c}{1+(h_d^2+h_c^2)P_p}\right),\\
\label{CRateConstraint12}
2R_c&\leq C\left(\frac{(h_d^2+h_c^2)P_c}{1+(h_d^2+h_c^2)P_p}\right),
\end{align}
where $P_p+P_c+\sum_{k=1}^KP_{cp}^{(k)}= P$, $P_{r}^{(1)}+P_{r}^{(2)}\leq P$, $R_{cp}= \sum_{k=1}^{K}R_{cp}^{(k)}$, $K\in\mathbb{N}$, and where the constraints \eqref{CPRateConstraint1}-\eqref{CPRateConstraint3} on the next page are satisfied.
\end{theorem}
\begin{figure*}
\begin{align}
\label{CPRateConstraint1}
R_{cp}^{(k)}&\leq C^+\left(\frac{h_{sr}^2P_{cp}^{(k)}}{1+2h_{sr}^2\sum_{i=k+1}^KP_{cp}^{(i)}+2h_{sr}^2P_c+2h_{sr}^2P_p}-\frac{1}{2}\right)&\forall k\in\{1,\dots,K\}\\
\label{CPRateConstraint2}
R_{cp}^{(k)}&\leq C\left(\frac{h_{d}^2P_{cp}^{(k)}}{1+h_{d}^2\sum_{i=k+1}^KP_{cp}^{(i)}+h_{c}^2\sum_{i=k}^KP_{cp}^{(i)}+h_{d}^2P_c+h_{c}^2P_c+h_{d}^2P_p+h_{c}^2P_p+h_r^2P_r^{(2)}}\right)&\forall k\in\{1,\dots,K\}\\
\label{CPRateConstraint3}
R_{cp}&\leq C\left(\frac{h_{r}^2P_{r}^{(1)}}{1+h_{r}^2P_{r}^{(2)}+h_{d}^2P+h_{c}^2P}\right)+C\left(\frac{h_{r}^2P_{r}^{(2)}}{1+h_{d}^2P_c+h_{c}^2P_c+h_{d}^2P_p+h_{c}^2P_p}\right).
\end{align}
\hrule
\end{figure*}

\begin{theorem}
\label{FDFSI}
The sum-rate $R_\Sigma= 2(R_c+R_{cp})$ is achievable where $R_{c}\leq \min\left\{C\left(\min\{h_d^2,h_c^2\}P_c\right),\frac{1}{2} C\left(h_d^2P_c+h_c^2P_c\right)\right\}$, and $R_{cp}$ satisfies \eqref{CPRateConstraintSI1}-\eqref{CPRateConstraintSI3} on the next page, such that $P_c+\sum_{k=1}^KP_{cp}^{(k)}= P$, $P_{r}^{(1)}+P_{r}^{(2)}\leq P$, $R_{cp}= \sum_{k=1}^{K}R_{cp}^{(k)}$, $K\in\mathbb{N}$.
\end{theorem}
\begin{figure*}
\begin{align}
\label{CPRateConstraintSI1}
R_{cp}^{(k)}&\leq C^+\left(\frac{h_{sr}^2P_{cp}^{(k)}}{1+2h_{sr}^2\sum_{i=k+1}^KP_{cp}^{(i)}+2h_{sr}^2P_c}-\frac{1}{2}\right)&\forall k\in\{1,\dots,K\}\\
\label{CPRateConstraintSI2}
R_{cp}^{(k)}&\leq C\left(\frac{h_{c}^2P_{cp}^{(k)}}{1+h_{c}^2\sum_{i=k+1}^KP_{cp}^{(i)}+h_{d}^2\sum_{i=k}^KP_{cp}^{(i)}+h_{d}^2P_c+h_{c}^2P_c+h_r^2P_r^{(2)}}\right)&\forall k\in\{1,\dots,K\}\\
\label{CPRateConstraintSI3}
R_{cp}&\leq C\left(\frac{h_{r}^2P_{r}^{(1)}}{1+h_{r}^2P_{r}^{(2)}+h_{d}^2P+h_{c}^2P}\right)+C\left(\frac{h_{r}^2P_{r}^{(2)}}{1+h_{d}^2P_c+h_{c}^2P_c}\right).
\end{align}
\hrule
\end{figure*}

\begin{proof}
Due to the lack of space, we refer the reader to \cite{NazerGastpar,ErezZamir,NarayananPravinSprintson} for more details about nested-lattice coding. In this work, we need nested-lattice codes with a fine lattice $\Lambda_f$ and a coarse lattice $\Lambda_c\subseteq\Lambda_f$ denoted $(\Lambda_f,\Lambda_c)$. The nested-lattice codewords are constructed as $x^n=(\lambda-d)\mod\Lambda_c$ where $\lambda\in\Lambda_f\cap\mathcal{V}(\Lambda_c)$ ($\mathcal{V}(.)$ for fundamental Voronoi region) and $d$ is a random dither. 

\subsection{Message splitting}
For a transmission block $b$, user 1 splits its message $m_1(b)$ into three parts, a private (P), a common (C), and a cooperative public (CP) \cite{PrabhakaranViswanath_SC} part denoted $m_{1,p}(b)$, $m_{1,c}(b)$, and $m_{1,cp}(b)$, respectively. Moreover, the CP message is divided into $K$ CP sub-messages $m_{1,cp}^{(k)}(b)$, $k=1,\dots,K$. The rates of these messages are denoted $R_{p},R_{c},R_{cp}^{(1)},R_{cp}^{(2)},\dots,R_{cp}^{(K)}$.

\subsection{Encoding}
Briefly, $m_{1,p}(b)$ and $m_{1,c}(b)$ are encoded into $x_{1,p}^n(b)$ and $x_{1,c}^n(b)$, respectively, where $X_{1,p}\sim\mathcal{N}(0,P_{p})$ and $X_{1,c}\sim\mathcal{N}(0,P_{c})$. Each CP message $m_{1,cp}^{(k)}(b)$ is encoded into $x_{1,cp}^{(k),n}(b)=(\lambda_{1,cp}^{(k)}(b)-d_{1,cp}^{(k)})\mod\Lambda_c^{(k)}$ using a nested-lattice code $(\Lambda_f^{(k)},\Lambda_c^{(k)})$ with power $P_{cp}^{(k)}$. In order to satisfy the power constraint, we set $P_p+P_{c}+\sum_{k=1}^KP_{cp}^{(k)}= P$. Same is done at transmitter 2, using the same nested-lattices. This enables the relay to decode the sum \cite{NarayananPravinSprintson} $u^{(k)}(b)=\lambda_{1,cp}^{(k)}(b)+\lambda_{2,cp}^{(k)}(b)$ modulo $\Lambda_c^{(k)}$. The transmitters then send the superposition of their codes as
\begin{align*}
x_j^n(b)=x_{j,p}^n(b)+x_{j,c}^n(b)+\sum x_{j,cp}^{(k),n}(b)\quad j\in\{1,2\},
\end{align*}
at each block $b\in\{1,\dots,B-1\}$. No messages are sent in block $B$. This incurs a rate loss which, however, becomes negligible for large $B$.

\subsection{Relay Processing}
In this scheme, the relay only decodes the CP messages. More precisely, the relay decodes the superposition of CP messages as follows. The relay starts decoding at the end of block $b=1$ where the sum $u^{(k)}(1)=(\lambda_{1,cp}^{(k)}(1)+\lambda_{2,cp}^{(k)}(1))\mod\Lambda_c^{(k)}$ is decoded, starting with $k=1$ and ending with $k=K$ (see successive compute-and-forward \cite{Nazer_IZS2012}). Decoding this superposition of codewords is possible as long as the rate constraint \eqref{CPRateConstraint1} is satisfied. 

Notice that the set of all possible values of $u^{(k)}(1)\in\mathcal{U}^{(k)}$ has size $\left|\mathcal{U}^{(k)}\right|=2^{nR_{cp}^{(k)}}$. The relay combines all $u^{(k)}(1)$, $k=1,\dots,K$, into one message $m_r\in\mathcal{M}_r$. Then the message set $\mathcal{M}_r$ has a size which is equal to the size of the Cartesian product of all $\mathcal{U}^{(k)}$, i.e., $\left|\mathcal{M}_r\right|=\left|\mathcal{U}^{(1)}\times\mathcal{U}^{(2)}\times\dots\times\mathcal{U}^{(K)}\right|=2^{n\sum_{k=1}^KR_{cp}^{(k)}}$. The relay then maps the message tuple $(u^{(1)}(1),\dots,u^{(K)}(1))$ to a message $m_r(2)\in\mathcal{M}_r$ to be sent in block $b=2$. This message is split into $m_r^{(1)}(2)$ and $m_r^{(2)}(2)$ with rates $R_{r}^{(1)}$ and $R_{r}^{(2)}$, respectively. The relay messages are then encoded to $x_r^{(1),n}(2)$ and $x_r^{(2),n}(2)$, two Gaussian codewords with powers $P_r^{(1)}$ and $P_r^{(2)}$, respectively, such that $P_r^{(1)}+P_r^{(2)}\leq P$. The sum of these codewords is sent in block $2$. This process is repeated for every block $b=1,\dots,B-1$. The relay sends in blocks $b=2,\dots,B$ and does not send any signal in block 1.

\subsection{Decoding}
\label{DecodingAtDestination}
The receivers wait until the end of block $B$ where decoding starts. Let us focus on receiver 1. At the end of block $B$ where only the relay is active, receiver 1 has $y_1^n(B)=h_r(x_r^{(1),n}(B)+x_r^{(2),n}(B))+z_1^n$ since the transmitters do not send in this block. Then, $m_r^{(1)}(B-1)$ and $m_r^{(2)}(B-1)$ are decoded successively in this order, which is reliable if
\begin{align}
\label{FDFDRSB1}
R_{r}^{(1)}&\leq C\left(\frac{h_r^2P_r^{(1)}}{1+h_r^2P_r^{(2)}}\right)\ \ \text{and}\ \ R_{r}^{(2)}\leq C(h_r^2P_r^{(2)}).
\end{align}

Now, the receiver knows $(u^{(1)}(B-1),\dots,u^{(K)}(B-1))$. Decoding proceeds backwards to block $B-1$ where
\begin{align*}
&y_1^n(B-1)\nonumber\\
&=h_dx_{1,p}^n(B-1)+h_dx_{1,c}^n(B-1)+h_d\sum x_{1,cp}^{(k),n}(B-1)\nonumber\\
&\quad+h_cx_{2,p}^n(B-1)+h_cx_{2,c}^n(B-1)+h_c\sum x_{2,cp}^{(k),n}(B-1)\nonumber\\
&\quad+h_rx_r^{(1),n}(B-1)+h_rx_r^{(2),n}(B-1)+z_1^n.
\end{align*}
The receiver decodes the messages successively in this order:
$m_{r}^{(1)}\to m_{1,cp}^{(1)}\to m_{1,cp}^{(2)}\to\dots\to m_{1,cp}^{(K)} \to m_r^{(2)}\to(m_{1,c},m_{2,c})\to m_{1,p}$. The message $m_{r}^{(1)}(B-1)$ is first decoded while treating the other signals as noise, leading to the first term in the rate constraint \eqref{CPRateConstraint3}. Next, the receiver decodes $m_{1,cp}^{(1)}(B-1)$ while treating the other signals as noise. Thus, we have the rate constraint in \eqref{CPRateConstraint2} with $k=1$.  

Recall that $u^{(1)}(B-1)$ is known at the receiver from the decoding process in block $B$. Now interference cancellation is performed. Since the receiver now knows both $m_{1,cp}^{(1)}(B-1)$ and $u^{(1)}(B-1)$, then, it can extract $m_{2,cp}^{(1)}(B-1)$ (see \cite{NarayananPravinSprintson}). It thus removes its contribution, $h_cx_{2,cp}^{(1),n}(B-1)$, from $y_1^n(B-1)$. Therefore, after decoding each $m_{1,cp}^{(k)}(B-1)$, interference from $m_{2,cp}^{(k)}(B-1)$ is cancelled. This continues until all CP messages are decoded, leading to the rate constraint \eqref{CPRateConstraint2}. At this stage, the receiver can calculate
\begin{align*}
&h_dx_{1,p}^n(B-1)+h_dx_{1,c}^n(B-1)+h_cx_{2,p}^n(B-1)\nonumber\\
&+h_cx_{2,c}^n(B-1)+h_rx_r^{(2),n}(B-1)+z_1^n.
\end{align*}
by subtracting the contribution of $m_r^{(1)}(B-1)$, $m_{1,cp}^{(k)}(B-1)$, and $m_{2,cp}^{(k)}(B-1)$, for $k=1,\dots,K$, from $y_1^n(B-1)$. The receiver then decodes $m_r^{(2)}(B-1)$,  $(m_{1,c}(B-1),m_{2,c}(B-1))$ (jointly), and $m_{1,p}(B-1)$ successively in this order, each time treating the remaining signals as noise. This leads to the second term in the rate constraint \eqref{CPRateConstraint3}, and the constraints \eqref{PRateConstraint}-\eqref{CRateConstraint12}. Notice that the first and second terms in \eqref{CPRateConstraint3} are more binding than \eqref{FDFDRSB1}, thus the latter are ignored. Additionally, since we have $R_r^{(1)}+R_r^{(2)}=R_r=\sum_{k=1}^KR_{cp}{(k)}=R_{cp}$, then, we can write the bound \eqref{CPRateConstraint3}. 

Decoding then proceeds backwards till block 1 is reached and the same is done at the second receiver, which proves the achievability of Theorem \ref{FDFWI}. Theorem \ref{FDFSI} can be proved similarly, except that the interfering CP messages are decoded first at each receiver instead of the desired CP messages. 
\end{proof}

At this point, it is worth to remark that a lattice strategy for the IRC was also proposed in \cite{TianYener_PotentRelayJournal}. The first difference between our scheme and the one in \cite{TianYener_PotentRelayJournal} is that we use P, C, and a set of CP messages, while in \cite{TianYener_PotentRelayJournal} each user sends only a CP message. The relay processing of the CP messages is the similar in both cases. The fundamental difference however is the decoding at the destination. We use interference cancellation described in Section \ref{DecodingAtDestination} which is not used in \cite{TianYener_PotentRelayJournal}.

To examine the performance of the FDF scheme, one has to carefully choose $K$ (the influence of which is explained in the next section) and the power allocations, plug in the FDF rate constraints, and compare to the upper bounds. In this way, it is possible to prove that the GDoF in Theorem \ref{Theorem:GDoF} is achievable. Due to space constraints, we use an example to illustrate the proof.

\section{GDoF: An Example}
\label{Section:GDoF}


Consider an IRC with $\beta-1<\gamma\leq\alpha\leq1\leq\beta$, and $2\alpha>1+\gamma$. In this case, from \eqref{GDoFThm} we obtain $d=\min\{2\alpha,1+\beta-\alpha\}$. Let us set the FDF parameters to $P_p=1/h_c^2$, $P_c=h_d^2P/h_r^2-P_p$, $P_r^{(1)}=P$, $P_r^{(2)}=0$,
$K=\left\lceil\log\left(h_{r}^2/h_d^2\right)/\log\left(h_d^2/h_c^2\right)\right\rceil$,
$P_{cp}^{(K)}=P\left(\frac{h_c^2}{h_d^2}\right)^{K-1}-P_c-P_p$, and 
$P_{cp}^{(k)}=P\left(\frac{h_c^2}{h_d^2}\right)^{k-1}-P\left(\frac{h_c^2}{h_d^2}\right)^{k},$
for $k=1,\dots,K-1$. Evaluating the expressions stated in Theorem \ref{FDFWI}, gives the achievable private GDoF $d_{p}= 1-\alpha$. For the common messages we get $d_{c}= \min\left\{2\alpha-\beta,(1+\alpha-\beta)/2\right\}$ where we used $\gamma>\beta-1$ and $2\alpha>1+\gamma$. For the cooperative public messages, by plugging the chosen parameters in \eqref{CPRateConstraint2} we get
\begin{align}
\label{dcpk}
d_{cp}^{(k)}&\leq1-\alpha,\quad\forall k=1,\dots,K-1,\\
\label{dcpK}
d_{cp}^{(K)}&\leq(K-1)(\alpha-1)-1+\beta.
\end{align} 
Here comes the importance of the choice of $P_{cp}^{(k)}$ and $K$. The choice of the powers of the CP signals leads to
$h_{d}^2P_{cp}^{(k+1)}=h_{c}^2P_{cp}^{(k)}$, i.e., while decoding the $k$-th CP message, the interference power from the ($k+1$)-th desired CP message is equal to that of the $k$th interfering CP message. Thus, the ($k+1$)-th desired CP message does not affect $d_{cp}^{(k)}$. This allows the achievability of $1-\alpha$. Now notice that without CP message splitting, that is all we could achieve. By splitting the CP messages, after decoding the $k$-th desired CP message, we can cancel the interference of the $k$-th interfering CP message, and then proceed to decode the ($k+1$)-th desired CP message where we have $h_{d}^2P_{cp}^{(k+2)}=h_{c}^2P_{cp}^{(k+1)}$, achieving another $1-\alpha$. By an appropriate choice of $K$, the first $K-1$ CP messages have $1-\alpha$ GDoF, leading to \eqref{dcpk}. While decoding the $K$-th desired CP message, the strongest interferer is the desired C message since $$h_{c}^2\left(h_c^2/h_d^2\right)^{K-1}P=h_{d}^2\left(h_c^2/h_d^2\right)^{K}P\leq h_{d}^4P/h_r^2,$$
which follows from the choice of $K$. In fact, $K$ is chosen as the largest number such that $K(1-\alpha)\geq\beta-1$ leading to the total CP GDoF $d_{cp}= \beta-1$. Interestingly, this is as if there were no CP interference at all, where $\beta-1$ would be achievable by decoding the CP messages while treating only the C and the P messages as noise. CP message splitting and interference cancellation therefore provide $d_{cp}= \beta-1$ instead of $d_{cp}=1-\alpha$. Similar CP GDoF expressions are obtained by evaluating the bounds \eqref{CPRateConstraint1} and \eqref{CPRateConstraint3}. Consequently, \eqref{CPRateConstraint1} and \eqref{CPRateConstraint3} do not decrease the achievable CP GDoF, which is still $\beta-1$. By adding $d_p$, $d_c$, and $d_{cp}$, we get the overall achievable GDoF of $d\leq \min\{2\alpha,1+\beta-\alpha\}$.

\begin{appendices}

\section{Proof of \eqref{Bound1} in Theorem \ref{SumBound2}}
\label{Proof:SumBound2}
The first bound in Theorem \ref{SumBound2}, i.e., \eqref{Bound1} is obtained by giving $Y_r^n$ and $(Y_r^n,m_1)$ as side information to receivers 1 and 2, respectively. Using classical information theoretic procedures, it is possible to write
\begin{align*}
n(R_\Sigma-\epsilon_n)&\leq I(m_1,m_2;Y_r^n)+h(h_dX_1^n+h_cX_2^n+Z_1^n|Y_r^n)\nonumber\\
&\quad-h(S_c^n|S_{sr}^n)+h(S_d^n|S_{sr}^n)-h(Z_2^n)
\end{align*}
with $\epsilon_n\to0$ as $n\to\infty$, $S_{sr}=h_{sr}X_2+Z_r$, $S_{c}=h_{c}X_2+Z_1$, and $S_{d}=h_{d}X_2+Z_2$. We proceed by writing
\begin{align*}
n(R_\Sigma-\epsilon_n)&\leq I(m_1,m_2;Y_r^n)+h(h_dX_1^n+h_cX_2^n+Z_1^n)\nonumber\\
&\quad-h\left(S_c^n/h_c|S_{sr}^n\right)+h\left(S_d^n/h_d|S_{sr}^n\right)\nonumber\\
&\quad+(n/2)\log\left(h_d^2/h_c^2\right)-h(Z_2^n),
\end{align*}
which follows since conditioning does not increase entropy, and since $h(aX)=h(X)+\frac{1}{2}\log(a^2)$. Now if $h_c^2\leq h_d^2$, then we can write $Z_1^n/h_c=\tilde{Z}^n+Z_2^n/h_d$, where $\tilde{Z}^n$ and $Z_2^n$ are independent. Then we can write
\begin{align*}
h\left(\frac{S_d^n}{h_d}\left|S_{sr}^n\right.\right)-h\left(\frac{S_c^n}{h_c}\left|S_{sr}^n\right.\right)=-I\left(\tilde{Z}^n;\frac{S_d^n}{h_d}+\tilde{Z}^n\left|S_{sr}^n\right.\right),
\end{align*}
which is negative. As a result, by letting $n\to\infty$, and using the Gaussian distribution for $X_1$ and $X_2$ to maximize the upper bound, we obtain \eqref{Bound1}. If $h_c^2>h_d^2$, then the bound \eqref{Bound1} can be obtained by enhancing receiver 2 by replacing the noise $Z_2$ by $h_dZ_2/h_c$, and proceeding as above.

\end{appendices}

\bibliography{myBib}		

\end{document}